
    \magnification 1200
    \hoffset .5 true in
    \hsize = 6.0  true in
    \vsize = 8.0 true in
    \parskip = 6pt          
    \font\tentworm=cmr10 scaled \magstep2
    \font\tentwobf=cmbx10 scaled \magstep2
    \font\tenonerm=cmr10 scaled \magstep1
    \font\tenonebf=cmbx10 scaled \magstep1
    \font\eightrm=cmr8
    \font\eightit=cmti8
    \font\eightbf=cmbx8
    \font\eightsl=cmsl8
    \font\sevensy=cmsy7
    \font\sevenm=cmmi7
    \font\twelverm=cmr12
    \font\twelvebf=cmbx12

    \def\subsection #1\par{\noindent {\bf #1} \noindent \rm}
    \def\mid {\let\rm=\tenonerm \let\bf=\tenonebf \rm \bf}
    \def\para{\par \vskip 12 pt}
    \def\head{\let\rm=\tentworm \let\bf=\tentwobf \rm \bf}
    \def\heading #1 #2\par{\centerline {\head #1} \smallskip
     \centerline {\head #2} \vskip .15 pt \rm}
    \def\eight{\let\rm=\eightrm \let\it=\eightit \let\bf=\eightbf
    \let\sl=\eightsl \let\sy=\sevensy \let\m=\sevenm \rm}

    \def\foots{\noindent \eight \baselineskip=10 true pt \noindent
\rm}
    \def\sexion{\let\rm=\twelverm \let\bf=\twelvebf \rm \bf}

    \def\section #1 #2\par{\vskip 20 pt \noindent {\mid #1} \enspace
{\mid #2}
    \para \noindent \rm}

    \def\ssection #1 #2\par{\noindent {\mid #1} \enspace {\mid #2}
    \para \noindent \rm}

    \def\abstract#1\par{\para \foots {\bf Abstract: \enspace}#1 \para}

    \def\author#1\par{\centerline {#1} \vskip 0.1 true in \rm}

    \def\abstract#1\par{\noindent {\bf Abstract: }#1 \vskip 0.5 true
in \rm}

    \def\midsection #1\par{\noindent {\sexion #1} \noindent \rm}

    \def\sqr#1#2{{\vcenter{\vbox{\hrule height#2pt
     \hbox {\vrule width#2pt height#1pt \kern#1pt
    \vrule width#2pt}
    \hrule height#2pt}}}}
    
    \def \dk {\vert\delta_k\vert}
    \def \dk2 {{\vert\delta_k\vert}^2}

\def \doublespace {\baselineskip = 20pt plus 7pt \message {double
space}}
\def \singlespace {\baselineskip = 15pt plus 3pt \message {single
space}}
\let \spacing = \doublespace
\doublespace

\def \body {\vfill \eject \parindent = 1.0 true cm
    \ifx \spacing \singlespace \singlespace \else \doublespace \fi}

\def\head{\let\rm=\tentworm \let\bf=\tentwobf \rm \bf}
\def\heading #1 #2\par{\centerline {\head #1} \smallskip
 \centerline {\head #2} \vskip .15 pt \rm}
\def \title#1 {\centerline {{\bf #1}}}
\def \Abstract#1 {\noindent \baselineskip=15pt plus 3pt \parshape=1
40pt310pt
  {\bf Abstract} \ \ #1}








\catcode`@=11
\def \C@ncel#1#2 {\ooalign {$\hfil#1 \mkern2mu/ \hfil $\crcr$#1#2$}}
\def \gf#1 {\mathrel {\mathpalette \c@ncel#1}}  
\def \Gf#1 {\mathrel {\mathpalette \C@ncel#1}}  

\def \gapx {\;\lower 2pt \hbox {$\buildrel > \over {\scriptstyle
{\sim}}$}
\; }
\def \lapx {\;\lower 2pt \hbox {$\buildrel < \over {\scriptstyle
{\sim}}$}
\; }


\def\n{\noindent}

\def\v2{\vskip 0.2 true cm}
\def\v3{\vskip 0.3 true cm}
\def\v4{\vskip 0.4 true cm}

\hyphenation {dimen-sional}


\heading{Getting the Measure of the Flatness Problem}

\vfill
\vskip 1.0cm
\centerline{By \bf{Guillaume Evrard$^{1,2}$} and {\bf Peter
Coles$^2$}}
\vskip 0.5cm
\centerline{$^1$ Groupe de Recherche en Astronomie et Astrophysique
du
Languedoc,}
\centerline{URA 1368 CNRS/Montpellier II, c.c. 072,
Universit\'e Montpellier II,}
\centerline{place Eug\`ene Bataillon, F-34095 Montpellier Cedex 05,
France}
\vskip .2cm
\centerline{$^2$ Astronomy Unit, School of Mathematical Sciences,}
\centerline{Queen Mary and Westfield College, Mile End Road,}
\centerline{London E1 4NS, United Kingdom}
\vfill
\bigskip
\vskip .5cm
\centerline{\bf Summary}
\vskip .4cm
{\bf The problem of estimating cosmological parameters
such as $\Omega$ from noisy or incomplete data is an
example of an inverse problem and, as such, generally
requires a probablistic approach. We adopt the Bayesian
interpretation of probability for such problems
and stress the connection between probability and information
which this approach makes explicit.
 This connection is important even when information
is ``minimal'' or, in other words, when we need to argue from
a state of maximum ignorance. We use the transformation group
method of Jaynes to assign minimally--informative prior
probability measure for cosmological parameters in the simple
example of a dust Friedman model, showing that the
usual statements of the cosmological flatness problem are based
on an inappropriate choice of prior. We further demonstrate that,
in the framework of
a classical cosmological model, there is no flatness problem.}

\bigskip
In the physical sciences, the word ``model'' is usually used to
denote a theoretical description of a system that contains one or
more ``free parameters'' whose values can not be determined {\it
a priori} but which have to be estimated by empirical means.
Such estimation problems generally go under the name of ``inverse
problems'' and, because available data are often incomplete or
noisy, they generally require probabilistic reasoning.

Modern `Big Bang' cosmology rests on a mathematical framework
supplied by the simplest relativistic cosmological models
compatible with the Cosmological Principle, i.e. the Friedman
models. These models have two free parameters, the Hubble
parameter, $H_0$, and the deceleration parameter $q_0$
(or, equivalently for these models, the deceleration parameter
$q_0=\Omega_0/2$; the suffix ``$0$''
indicates that the parameter in question is measured
at the present epoch, i.e. when the cosmological proper time
is $t_0$.) As
is the case for physical models in general, these parameters are not
{\it predicted} by the Big Bang theory itself, but need to be
inferred from observational data. Because the values of $H$
and $\Omega$ at any time can be determined from the present
values $H_0$ and $\Omega_0$ if the model is specified, it
is in principle possible to learn about conditions very
near the Big Bang singularity from estimates of the
cosmological parameters made at the present time.

The problem with $\Omega$ is that its value is not known
with any precision: it probably lies in the range $0.10 < \Omega_0
<1.5$, but the relevant evidence is often contradictory$^1$.
However, $\Omega$ evolves strongly
with cosmic time $t$ in such a way that
$\Omega=1$ is an unstable
fixed point. To get a value of $\Omega$ anywhere near unity at the
present time (even a factor
of a few either way) consequently requires a value at very early
times extremely close to unity (say $\Omega=1\pm 10^{-60}$
at the Planck time). The {\it cosmological flatness problem}
arises from the judgement that this ``fine--tuning'' is somehow
unlikely on the basis of standard Friedman models; it is
is usually ``resolved'' by appealing to some transient mechanism
(e.g. inflation$^{2}$) which can make $\Omega$ evolve towards unity
for some time, rather than away from it.

 But do we have
any right to claim that some values of $\Omega$ are more likely than
others? Can one make any inferences at all from the
uncertain parameter estimates we have in cosmology? And
what precisely does it mean  to say that $\Omega$ is ``close to
unity''
anyway?

To answer these questions we need to understand the role of
probability
in the solution of inverse problems generally$^3$. We adopt the
objective Bayesian
interpretation of probability which, we believe, is the
only way to formulate this type of reasoning in a fully self--
consistent
way. In this interpretation, probability
represents a generalisation of the notions of ``true'' and ``false''
to intermediate cases where there is insufficient information to
decide with logical certainty between these two alternatives$^{4}$.
Unlike the opposing ``frequentist'' view, the Bayesian lends
itself naturally to the interpretation of unique events, of
which the Big Bang is the most obvious relevant example$^{5}$.

The central principle involved in Bayesian inference is Bayes'
theorem$^{6}$.
Suppose $H_i$ represents one of a set of hypotheses (or models),
$D$ is some data and $I$ is whatever relevant prior information we
may have
(or which we assume to be the case) before obtaining the data $D$.
Bayes' theorem states that
$$
P(H_i|DI)={P(H_i|I) P(D|H_iI)\over \sum_i P(H_i|I)P(D|H_i I)},
\eqno(1)
$$
where $P(H_i|I)$ is called the prior probability of $H_i$ given our
prior information, $P(D|H_iI)$ is the likelihood and $P(H_i|DI)$
is the posterior probability. Notice that all probabilities
here are conditional on the information $I$ which is either known
or assumed to be true in a given model. If the prior is relatively
flat and the likelihood of the data $D$ is strongly peaked for
a particular $H_i$ then our inference of the posterior probability
is strongly determined by the data. If, on the other hand, the data
discriminate only weakly between the models then the posterior
is dominated by the prior. In general, however, both prior
and likelihood are required for the inverse problem to be well--posed.
 Many critics have dubbed the Bayesian approach
``subjective'' because different individuals may
possess different information and therefore assign different
priors to the same hypothesis. This is not a serious objection:
your assessment of the probability that a given horse will win
a race must change if you learn the other horses have all been
drugged! What is important is that, given the same information,
the same prior should be assigned. We therefore need an
objective set of rules
for assigning priors when  information is specified.
In particular, we may have no information at all
other than that inherent in the model we adopt.
What  should one do when one has such
minimal information about a system?

Even this apparently simple question
turns out to be extremely deep and there is no universally
accepted principle for assigning minimally--informative priors
in general circumstances. Jaynes$^{7}$ has described one
approach which is, as far as we are aware, the most general
objective algorithm available. ``Jaynes' principle''
is that one looks for a measure on the parameter space
of the system that possesses the property of invariance
under the group of transformations which leave
unchanged the mathematical form of the physical laws describing
the system. In the absence of any other constraints, the
principle of maximum information entropy (a principle of
least prejudice) yields a prior probability simply
proportional to this measure.

To take a trivial illustrative example, consider the problem of
estimating
the position of a particle on the real line. Our state of
knowledge, if no signposts are visible, must be  unchanged
if we shift our coordinates by any distance $\gamma$. This requires
$\mu(x)=\mu(x+\gamma)$, a functional equation which has only one
solution: $\mu=$constant. This is in full accord with our intuition,
but it does not mean that a uniform prior is appropriate for
all cases where we are seeking to encode minimal information.
For example, Evrard$^{8}$ has calculated the least--informative
prior for a free particle in velocity space using Jaynes' principle
and
the laws of special relativity. Even in this simple example, the
result is non--trivial:
 ``least information prior'' does  not necessarily mean ``no prior''.

We now turn to the appropriate
minimally informative prior for the cosmological parameters
$H_0$ and $\Omega_0$. We take the laws of
physics to be the Friedman equations
describing a pressureless perfect fluid in the form
$$
a\Bigl(k+{\dot{a}^{2}\over c^{2}}\Bigr) =\chi, \eqno(2)
$$
where $\chi$ remains constant throughout the evolution of the
system; its value is determined by the ``initial value
equation''
$$
\chi={4\pi G \rho a^{3}\over 3c^{2}}. \eqno(3)
$$
The quantity $\chi$ can be thought of as an absolute scale parameter.
In equations (2) \& (3), $a$ is the cosmic scale factor
(another scale parameter) and $\rho$ is the
matter density. The quantity $k$ appearing in equation (2)
is the
curvature of spatial
sections in the model, scaled to take the values $0$
if $\Omega=1$, $-1$ if $\Omega<1$ or $+1$ if $\Omega_0>1$.
 The system can be parametrised
completely in terms of $\chi$ and $a$.
(In fact,  we could equally well have chosen to work with redshift
$z$, cosmological proper time $t$, conformal time $\tau$, temperature
$T$, or anything else monotonically related to $a$: the resulting
measure would turn out to be the same, but the equations
turn out to be simpler in terms of $a$ itself.) We now need to
express the cosmological parameters $H=\dot{a}/a$ and $\Omega = 2q=-
2a\ddot{a}/\dot{a}^{2}$ in terms of $a$ and $\chi$.  We obtain,
for $k=\pm 1$,
$$
\Omega=2(2\mp a/\chi)^{-1}\eqno(4)
$$
and
$$
H=\Bigl({c\over \chi}\Bigr) {\sqrt{2\mp
a/\chi}\over(a/\chi)^{3/2}}.\eqno(5)
$$
Remember that the suffix $0$ represents a quantity defined at the
present epoch, so $H_0$ and $\Omega_0$ are the values
of these parameters  when $a=a_0$;
$\chi=\chi_0$ at all epochs. Because both  $\chi$ and
$a$ are scale parameters, we look for a measure which is invariant
under the transformations $a'=\alpha a$ and
$\chi'=\beta\chi$, where $\alpha$ and $\beta$ are constants.
Such invariances require that the information represented by our
measure does not change if we use a different ruler to measure
distances. It follows that
$$
\mu(\chi,a)\propto {1\over \chi a},\eqno(6)
$$
which becomes, after substituting from equations (4) \& (5),
$$
\mu(H,\Omega) \propto {1\over H\Omega|\Omega-1|}. \eqno(7).
$$
 Note that this measure
leads to an {\it improper} (i.e. non--normalisable) prior probability.
This can be rectified by bringing in additional information,
such as the ages of cosmic
objects which rule out high values of both $\Omega$ and $H$.
Anthropic selection effects can also be brought to bear on
this question$^5$. The measure for $H$ is uniform
in the logarithm, as one might expect from the Bayesian ``rule of
thumb''
for scale parameters$^9$. The measure
in $\Omega$ is, however, more complicated than this. In particular,
it diverges at $\Omega=0$ and $\Omega=1$, the former corresponding to
an empty Universe without deceleration and the latter to the
critical-density Einstein--De Sitter model.
These singularities
could have been anticipated because these are two fixed points
in the evolution of $\Omega$. A model with $\Omega=1$ exactly remains
in that
state forever. Models with $\Omega<1$ evolve to a state of free
expansion with $\Omega=q=0$. Since states with $0<\Omega<1$ are
transitory,
it is reasonable, in the absence of any other information, to infer
that the system should be in one of the two fixed states.
(All values of $\Omega>1$ are transitory.)

The measure (7) also demonstrates how dangerous it is to talk
about $\Omega_0$ ``near'' unity.
In terms of our least--informative measure, values of $\Omega$ not
exactly equal to $1$ are actually infinitely far from this value.
A similar property is held by the velocity--space measure$^{8}$,
which demonstrates the velocities of all material particles
are, in a well--defined sense, infinitely far from $c$.

We now turn to the flatness problem. The usual
argument is essentially that, without inflation, the models that
produce $\Omega_0=1\pm \epsilon$ at the present epoch emerge from
earlier states with $\Omega$ even closer to unity. If one were to
adopt a measure which is roughly flat in the vicinity of $\Omega=1$ as
$t\rightarrow 0$ then the
probability associated with this set of states
would vanish and there
would indeed be a flatness problem: it would appear
``unlikely'' that our Universe was correctly modelled by the
standard Friedman equations and one would be pushed into accepting
inflation as a solution of this ``fine--tuning''.
But our measure (7) demonstrates
that the assumption of a constant prior for $\Omega$
is not consistent with the assumption of minimal
information. It therefore represents
a considerable prejudice compared to the least---informative
and, therefore, least--prejudiced measure. This prejudice
may be motivated to some extent by quantum--gravitational
considerations that render the classical model
inappropriate, but unless the model adopted and its associated
information are stated explicitly one has no right to assign
a prior and therefore no right to make any inferences.

Notwithstanding the recent research interest in quantum gravity,
we feel that `minimal knowledge' is a fair description of our
state of understanding of physics at the Planck epoch.
In terms of the least--informative
measure, the probability associated with smaller and smaller
intervals of $\Omega$ (around unity) at earlier and earlier times
need
not become arbitrarily small because of the singularity at $\Omega=1$.
Indeed, this measure  is constructed in precisely such a way that
the probability associated
with a given range of $\Omega_0$ is
preserved as the system evolves. We should not therefore be
surprised to find $\Omega_0\simeq 1$ at the present epoch even
 in the absence of inflation, so
we do not need inflation to ``explain'' this value.
In this sense, {\it there is no flatness problem} in
a purely classical cosmological model.

We realise that many of the issues we have discussed remain
controversial.
We accept, for example,
that Jaynes' principle may be the last word in the theory
of prior assignment based on minimal information.
Nevertheless, inferences based only on vague prescriptions of uniform
priors have no place in physics
or cosmology. Consistent inverse reasoning {\it requires}
the assignment of a prior according to some objective rules;
failure to do this replaces bona fide inductive logic
with mere superstition.

\bigskip
\vskip 0.5cm
\centerline {Acknowledgments}
Peter Coles receives a PPARC Advanced Fellowship. Guillaume Evrard
acknowledges support from the European Community Human Capital \&
Mobility Programme (contract number ERBCHRX-CT93-0129) while this
paper
was written. A previous version of this paper received an honourable
mention in the Gravity Research Foundation Essay Competition 1995.

\vfill
\eject
\vskip .5cm
\centerline {\bf REFERENCES}
\vskip .4cm
\n
(1) P. Coles \& G.F.R. Ellis, 1994. {\it Nature}, {\bf 370}, 609--615.
\vskip .2cm
\n
(2) A.H. Guth, 1981. {\it Phys. Rev. D.}, {\bf 23}, 347--356.
\vskip .2cm
\n
(3) A. Tarantola \& B.J. Valette, 1982. {\it J. Geophys.}, {\bf 50},
159--170.
\vskip .2cm
\n
(4) R.T. Cox, 1946. {\it Am. J. Phys.}, {\bf 14}, 1--13.
\vskip .2cm
\n
(5) A.J.M. Garrett \& P. Coles, 1993. {\it Comments on Astrophys.},
{\bf 17}, 23--47.
\vskip .2cm
\n
(6) T.J. Loredo, 1990. In {\it Maximum Entropy and Bayesian Methods},
ed. P.F. Foug\`{e}re, pp. 81--142, Kluwer, Dordrecht.
\vskip .2cm
\n
(7) E.T. Jaynes, 1968. {\it IEEE Transactions on Systems Science and
Cybernetics}, {\bf SSC--4}, 227--241.
\vskip .2cm
\n
(8) G. Evrard, 1995, {\it Physics Letters}, {\bf A201}, 95--102.
\vskip .2cm
\n
(9) H. Jeffreys, 1939. {\it Theory of Probability}, Clarendon Press,
Oxford.

\vfill\eject
\bye